\documentclass[aps,pre,reprint,superscriptaddress,showpacs]{revtex4-1}
\usepackage{graphicx}
\usepackage{graphics}
\usepackage{amsmath}
\usepackage{amssymb}
\usepackage{amsfonts}
\usepackage{epsfig}

\begin{document}


\title{Different types of critical behavior in conservatively coupled H\'enon maps}


\author{Dmitry V. Savin}
\email[Corresponding author: ]{savin.dmitry.v@gmail.com}
\affiliation{Department of Nonlinear Processes, Chernyshevsky Saratov State University, Astrakhanskaya str.\,83, 410012, Saratov, Russia}
\author{Alexander P. Kuznetsov}
\affiliation{Department of Nonlinear Processes, Chernyshevsky Saratov State University, Astrakhanskaya str.\,83, 410012, Saratov, Russia}
\affiliation{Kotel'nikov Institute of Radioengineering and Electronics of RAS, Saratov branch, Zelenaya str.\,38, 410019, Saratov, Russia}
\author{Alexey V. Savin}
\affiliation{Department of Nonlinear Processes, Chernyshevsky Saratov State University, Astrakhanskaya str.\,83, 410012, Saratov, Russia}
\author{Ulrike Feudel}
\affiliation{Institute for Chemistry and Biology of the Marine Environment,
Carl von Ossietzky University Oldenburg, Carl von Ossietzky str.\,9-11, D-26111, Oldenburg, Germany}



\begin{abstract}
We study the dynamics of two conservatively coupled H\'enon maps at different levels of dissipation. It is shown that the decrease of dissipation leads to changes in the structure of parameter plane and the scenarios of transition to chaos compared to the case of infinitely strong dissipation. Particularly, the Feigenbaum line becomes divided into several fragments. Some of these fragments have critical points of different types, namely of C and H type, as their terminal points. Also the mechanisms of formation of these Feigenbaum line ruptures are described.
\end{abstract}


\maketitle

\section{\label{I}Introduction}
Nonlinear dynamical systems show a rich variety of different dynamical regimes such as stationary points, periodic and quasiperiodic orbits as well as chaotic motion depending on the system parameters. Bifurcations related to transitions between these dynamical regimes when a system parameter is varied have been studied extensively during the last three decades (see e.g. \cite{guck83, kuzyubif} and refs. therein). These studies include the well-known routes to chaos via period doubling cascades, quasiperiodicity and intermittency (\cite{argyris1994, alligood1997, ott2002, mann2010} and refs. therein). One of the striking features of chaos is that chaotic parameter regions are interrupted by periodic motion of different periods appearing in periodic windows \cite{milnor88, graczyk, joglekar}. Such close neighborhood of stable predictable motion on periodic orbits and irregular chaotic motion makes the system dynamics, particularly in real world applications, rather complex, since already slight perturbations in parameters can shift the system from periodic to chaotic motion and vice versa. While these windows emerge in many applications in different disciplines of science (see \cite{bonatto2008} for examples from laser physics, atmospheric science and chemistry) leading to quite intricate intertwined structures in parameter space, such windows are absent in systems possessing robust chaotic attractors, both hyperbolic (\cite{kuzbook} and refs. therein) and pseudo-hyperbolic \cite{turshi98}, particularly Lorenz-like \cite{afrBykov, guckLorenz, gonch3dhenon}. The fine structure of periodic windows has been investigated for paradigmatic maps such as the two-parameter quadratic map \cite{barreto97} and the H\'enon map \cite{gallashenon, lorenzhenon}. In the two-dimensional parameter space such windows  have a typical form which depends on the special type of organization of bifurcation lines for its main period. The two mostly common types are spring area and crossroad area structures \cite{Mira1, Mira2, Mira3} (periodic windows based on the crossroad area are often called shrimps \cite{gallashenon, stoop2012}). Such structures in parameter plane have been found in driven, parametrically excited and impact oscillators \cite{Medeiros2011, Zhou2006, Medeiros2010, oliveira2011shrimp}, electrochemical oscillators \cite{GallasChem}, two-gene systems \cite{Souza2012}, lasers \cite{GallasLaser, bonatto2007accumulation}, population dynamical systems in ecology as well as paradigmatic models such as R\"ossler system \cite{BarrioRossl}. These shrimp structures have been demonstrated to be observable in a hardware realization using electronic circuits \cite{Stoop2010}.

Another important feature of chaotic dynamical systems is the existence of different types of critical behavior on the border of chaos. If the transition to chaos occurs via the cascade of period-doubling bifurcations, the scaling properties on the border of chaos are in general determined by the Feigenbaum scaling law \cite{Feig1, Feig2}. Such situation is typical for one-dimensional maps. When the phase space as well as the parameter space are high-dimensional, the transition to chaos via the Feigenbaum scenario is also a common situation, and the border of the chaotic region is formed by the Feigenbaum critical surface to which period-doubling bifurcations accumulate. This surface could be bounded by some other surfaces (or lines) with smaller dimension. It turns out that the structure of the parameter space in the vicinity of these borders of the Feigenbaum critical surface and the structure of the phase space on this border has scaling properties which differ from the Feigenbaum scaling law (\cite{KuzStat} and refs. therein). In terms of the renormalization group analysis such critical behavior is associated with saddle points of the generalized Feigenbaum-Cvitanovi\'c equation, and the number of its eigenvalues with modulus greater than 1 determines the codimension of the critical point \cite{Kuz1992}. It turns out that in many cases it is much simpler to observe these critical points with high codimension in unidirectionally or mutually coupled systems with period-doublings (\cite{KuzStat} and refs. therein).

The dynamics of nonlinear systems depends in general not only on the nonlinearity but also on the level of dissipation. There are two main phenomena which characterize the system dynamics with the change of dissipation. First, while for strong dissipation the dynamics is characterized by only one attractor, the coexistence of a multitude of attractors for a given set of parameters is the norm for weakly dissipative systems \cite{Feudel1996}. As the conservative limit is approached more and more coexisting attractors appear as it has been shown for several model systems as the standard map \cite{Feudel1996, GallasMult}, the H\'enon map \cite{Feudel2003, RechBas} and in more realistic systems like a suspension bridge model \cite{Bridge} (see also the review in \cite{FeudelRev}). Second, of particular interest is the crossover from dissipative to conservative dynamics \cite{Zisook, Reinout, Chen1986}. This crossover is related to different scaling relations, e.g. for the accumulation of period doublings at the transition to chaos.

Investigations on weakly dissipative systems are usually carried out for a low-dimensional map or a system of ordinary differential equations which is autonomous or periodically driven. The effect of changes of dissipation on coupled systems has been only rarely studied, to our knowledge only the double kicked rotor has been analyzed to reveal the extremely high degree of multistability associated with complexly interwoven basins of attraction \cite{FeudelRev}.

In this paper we try to relate different of the aforementioned aspects of dynamical systems. We study coupled systems changing the dissipation level in order to reveal their different routes into chaos. In contrast to previous studies of the bifurcation scenarios in the two parameter space spanned by nonlinearity and coupling, we focus on a coupling which is fixed and conservative, i.e. it does not contribute to an increase in dissipation. Instead we analyze the dynamics in the two parameter space spanned by the two nonlinearity parameters of the two coupled systems following the approach developed in \cite{YuanLog, Satoh, KuzLog}. Each of the systems, when uncoupled, exhibits the period doubling route to chaos. When coupled, period doublings still occur, but also the quasiperiodicity route to chaos is observed. Our main aim is to study the critical behavior associated with the line of Feigenbaum accumulation points in parameter space. Changing the level of dissipation this Feigenbaum line ruptures and one finds several pieces of that line with different critical behavior at its ends. We show that this rupture occurs due to the movement of some periodic window in parameter space when the dissipation is varied. In fact, the rupture emerges as a result of a ``collision'' of this periodic window with the main periodic area possessing the transition to chaos via the continuous Feigenbaum line in parameter space. Further decrease of dissipation results in the appearance of another rupture of the Feigenbaum line caused by a sequence of Neimark-Sacker bifurcations for cycles from the period-doubling cascade.

The paper is organized as follows: in Section\,\ref{2} we present the model system under investigation, an overview of the parameter plane evolution and  details of the bifurcation structure of the parameter plane and the critical behavior at the border of chaos, in Section\,\ref{3} we analyze the mechanism of the rupture of the Feigenbaum line in more detail, and finally in Section\,\ref{4} we summarize the obtained results.

\section{\label{2}Two conservatively coupled H\'enon maps}

\subsection{The model system}

To study the dynamics in coupled dissipative systems we focus on two coupled H\'enon maps. Choosing these maps as the simplest paradigms for invertible maps has several advantages: (i) All results obtained should also occur when studying models represented by differential equations. (ii) The single H\'enon map, first introduced in \cite{henon1976two}, is one of the mostly studied maps over many years and has been previously employed to investigate different aspects of critical behavior on the border of chaos. Though the map itself was constructed as a paradigmatic theoretical model, it has also been used to describe experimental results, particularly for the description of the critical behavior in the Rayleigh-B{\'e}nard experiment \cite{arneodo1983observation}. (iii) In the limit of infinite dissipation we obtain two coupled quadratic maps, which have also been widely studied. In mathematical terms our model system can be written as:
\begin{equation}\label{maineq}
\begin{split}
x_{n+1} & =\lambda_1-{x_n}^2-b y_n+\varepsilon(x_n-u_n),\,y_{n+1}=x_n, \\
u_{n+1} & =\lambda_2-{u_n}^2-b v_n+\varepsilon(u_n-x_n),\,v_{n+1}=u_n.
\end{split}
\end{equation}
Here $\lambda_1$ and $\lambda_2$ are the forcing parameters, responsible for the emergence of the period doubling cascade in the single H\'enon map and $b$ is the damping parameter characterizing the level of dissipation. We use a linear diffusive coupling in the first variable with coupling strength $\varepsilon$.  This coupling is quite convenient since it will not introduce any extra dissipation in the system. Hence, in this case it is quite simple to control the dissipation level which is given by the Jacobian of the map $J=b^2$. In the limit $b=1$ the system is conservative, while in the limit $b=0$ the dissipation is infinite and Eq.\,(\ref{maineq}) transforms into two coupled quadratic maps. This allows us to vary the dissipation continuously in the interval $b\,[0,1]$ having well defined limits on both ends of the interval.
\subsection{Evolution of the parameter plane with decreasing dissipation}
\begin{figure}
 \includegraphics[width=8.6cm]{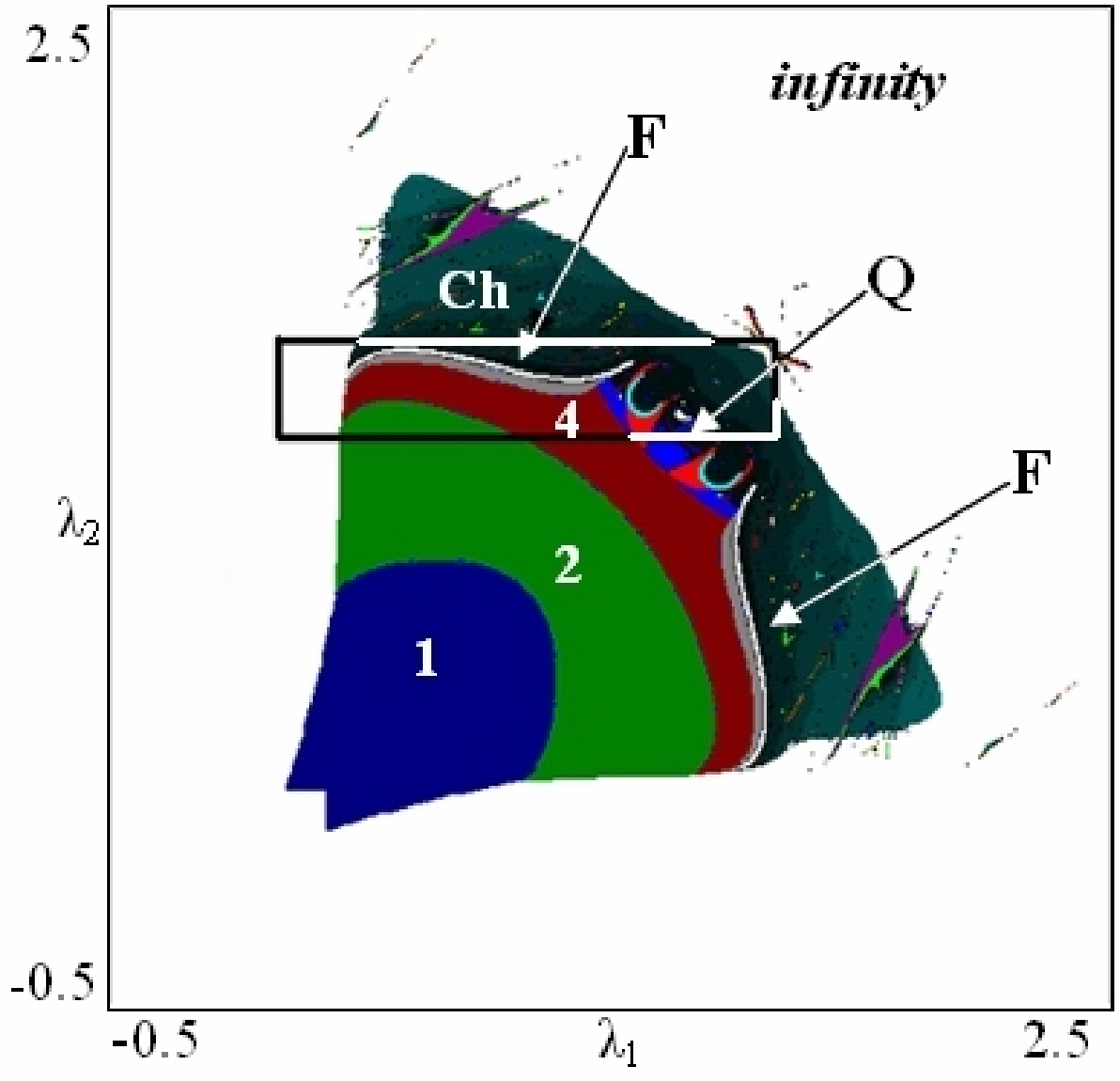}%
 \caption{\label{fig1}(color online) Structure of the parameter plane of Eq.\,(\ref{maineq}) at $b=0, \varepsilon=0.4$. The white line F denote the transition to chaos via the Feigenbaum scenario. Areas of different colors correspond to regions of the existence of cycles possessing certain period (see numbers in the figure), Ch denotes the area with a chaotic dynamics corresponding to positive larger Lyapunov exponent (chaotic regime), Q denotes area in which Lyapunov exponent is close to 0 (quasiperiodical regime), in the white region the trajectories go to infinity. The sketch of the Feigenbaum line in the dark rectangle is shown in Fig.\,{\ref{fig3}}.}
 \end{figure}

In the case of infinite dissipation Eq.\,(\ref{maineq}) turns into the system of linearly coupled logistic maps. The dynamics of the latter is investigated rather well \cite{YuanLog, Satoh, KuzLog}. The structure of the $(\lambda_1,\lambda_2)$ parameter plane in this case is shown in Fig.\,{\ref{fig1}}. It is known that besides the transition to chaos via the Feigenbaum period-doubling cascade the transition to chaos via quasiperiodicity can also be found in the region close to the diagonal of the parameter plane. Thereby, there exist two regions in the parameter plane with different scenarios of the transition to chaos (let us denote them as F and Q, respectively). On the border of these two regions the Feigenbaum line of the transition to chaos terminates in a codimension-2 critical point associated with the cycle of period 2 in the renormalization group equation \cite{KuzLog}. Such point is usually called critical point of C-type \cite{Kuz1992}.

\begin{figure*}
 \includegraphics[width=17.8cm]{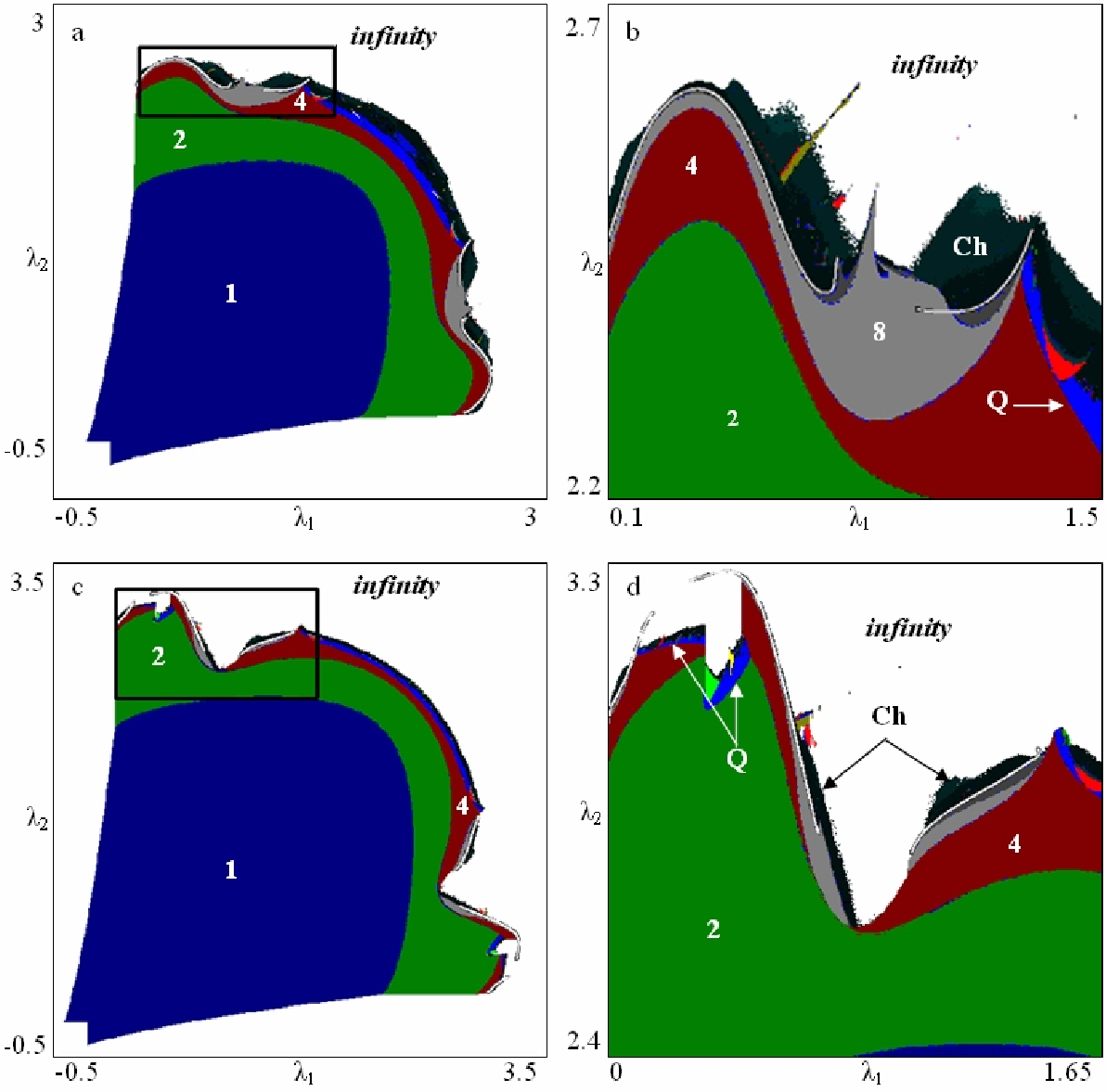}%
 \caption{\label{fig2}(color online) Structure of the parameter plane (a, c) and its enlarged fragments (b, d) of (3) at $\varepsilon=0.4$; $b=0.5$ (a, b), $b=0.7$ (c, d). Color code and symbols as in Fig.\,{\ref{fig1}}, black rectangles in the left column are enlarged in the right one. Fragments of the Feigenbaum line F extending into the ``infinity" region correspond to the attractor with a small basin of attraction which coexists with an attractor at infinity in the corresponding area of the parameter plane.}
 \end{figure*}

Let us now decrease the dissipation level by increasing the parameter $b$ towards 1. The structure of the parameter plane at different $b$ values is shown in Fig.\,{\ref{fig2}}. One can see that both F and Q regions change their structure with the decrease of dissipation. The quasiperiodicity area becomes much thiner but spreads into the region far from the diagonal of the parameter plane; in Figs.\,{\ref{fig2}}\,b,\,d the Neimark-Sacker (NS) bifurcation line, marking the border between the period-4 and quasiperiodicity areas, can be seen in the parameter region where  $\lambda_1$ and  $\lambda_2$ are sufficiently different. This situation contrasts to the case $b=0$ where the quasiperiodicity region exists only near the diagonal  \cite{YuanLog, Satoh, KuzLog} as in Fig.\,{\ref{fig1}}. The Feigenbaum line undergoes a rupture, and instead of one line on each side of the diagonal we observe at $b=0.5$ 2 pieces (Fig.\,{\ref{fig2}}\,a,\,b). Right after the rupture multistability with coexisting attractors appears: this can be seen in Fig.\,{\ref{fig2}}\,b in which the right fragment of the Feigenbaum line continues into the area of the period-8 cycle, which means that in this region of parameter space the stable period-8 cycle coexists with the sequence of period-doublings accumulating at the Feigenbaum line. The dynamics in between the two large pieces of Feigenbaum lines the dynamics is even more involved, but this complicated structure is beyond the scope of this paper.With further increase of $b$ ($b=0.7$ in Fig.\,{\ref{fig2}}\,c,\,d) a second Q area arises far from the diagonal of the parameter plane, which causes an additional rupture of the Feigenbaum line possessing now 3 pieces on each side of the diagonal. From these observations two questions arise:
\begin{enumerate}
\item How does the bifurcation structure around the terminal points of the fragments of the Feigenbaum line look like?
\item What is the mechanism of the rupture?
\end{enumerate}

To answer these questions we study the bifurcation structure in more detail using the continuation software CONTENT \cite{content} and investigate the critical behavior. It is worth mentioning here that the Feigenbaum lines shown in the figures are obtained as the limit of the subsequent period-doubling lines using the software CONTENT.

When the $b$ value is rather small, the Feigenbaum line is still continuous. In this case period-doubling (PD) lines of consecutive periods from the period-doubling cascade terminate in ``fold-flip'' points in which one pair of multipliers have modulus less than 1 while the other two are equal to (-1, +1) \cite{kuzyubif}. The coordinates of these points for the right end of the Feigenbaum line at $b=0.3$ are shown in Table\,\ref{tab1}. This sequence converges to a certain limit, which is the terminal point of the Feigenbaum line. A similar sequence could be observed for the second terminal point as well. Such sequences are known to have the critical points of C-type as their limit \cite{KuzStat}. We can conclude that at $b=0.3$ the Feigenbaum line terminates in two C-type critical points, as in two coupled logistic maps at $b=0$ \cite{KuzLog}.
 \begin{table}
 \caption{\label{tab1}Coordinates of the ``fold-flip'' points at the right end of the Feigenbaum line at $b=0.3$}
 \begin{tabular}{ccc}
\hline \hline
Period & $\lambda_1$ & $\lambda_2$ \\
\hline
8 & 1.19367393 & 2.02423176 \\
16 & 1.23587229 & 2.07183506\\
32 & 1.20996171 & 2.04250455\\
64 & 1.22555098 & 2.05983479\\
128 & 1.21947400 & 2.05297370\\
256 & 1.22371398 & 2.05774598\\
512 & 1.22252007 & 2.05639539\\
1024 & 1.22335567	 & 2.05734009\\
\hline \hline
 \end{tabular}
 \end{table}
With increasing $b$ the first rupture occurs at $0.3218<b<0.3219$, and at $b=0.5$ (Fig.\,{\ref{fig2}}\,a,\, b) the Feigenbaum line is already divided into two fragments in each half of the symmetric parameter plane. The terminal points of the PD lines are again ``fold-flip'' points, and these sequences also converge to C-type critical points.

\begin{figure}
 \includegraphics[width=8.6cm]{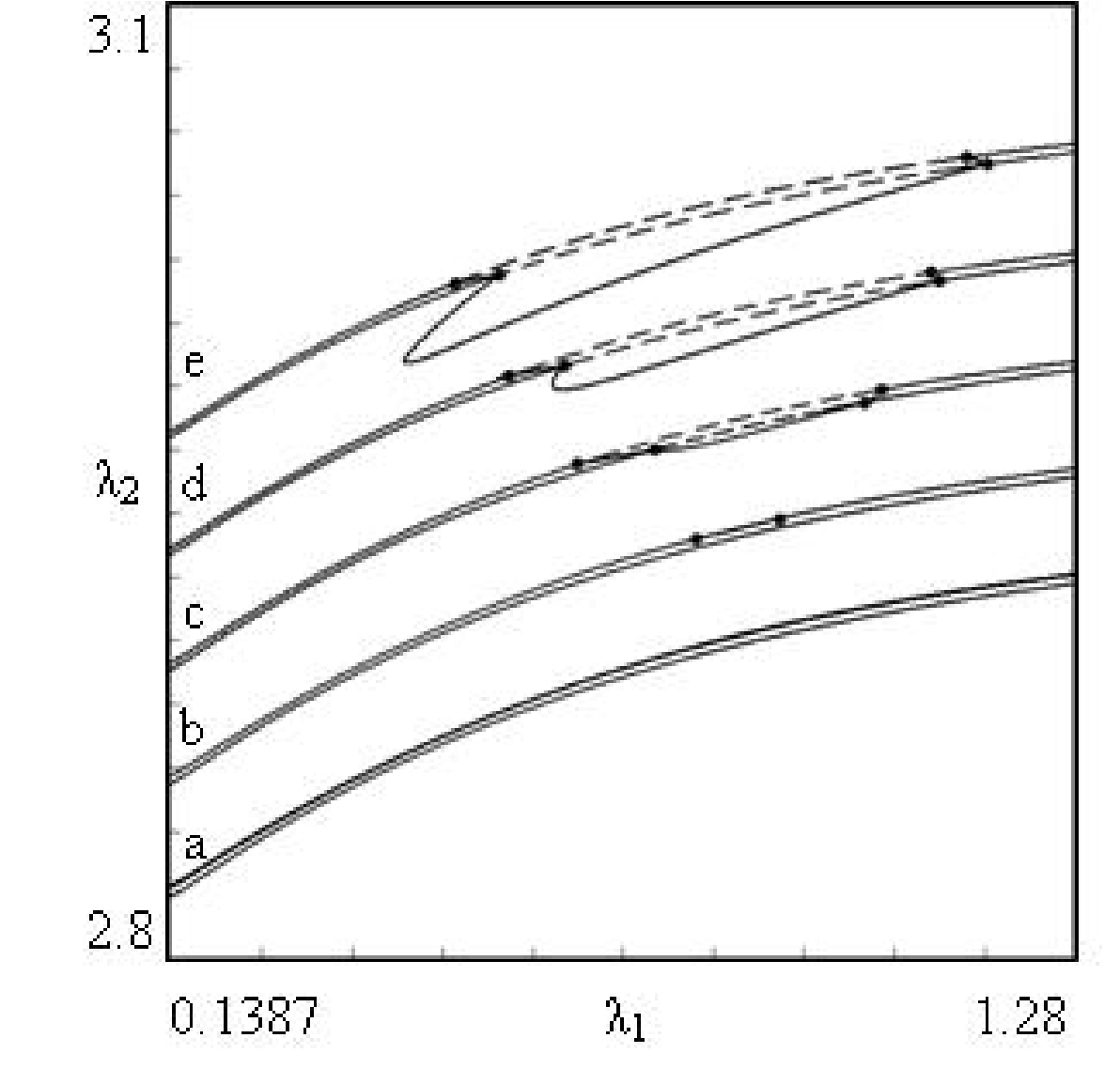}%
 \caption{\label{fignew} Bifurcation lines of  Eq.\,(\ref{maineq}) at increasing values of $b$, illustrating the process of formation of the quasiperiodicity area. Points denote resonance 1:2 points, solid lines --- lines of  period-doubling (PD) (outside of a pair of resonance 1:2 points) and Neimark-Sacker bifurcations (inside of resonance 1:2 points), dashed lines --- fragments of PD lines, corresponding to the unstable cycle. Each system (a-e) consists of two lines, bounding the stability region of cycle of period 8 (low line) and 16 (upper line). Values of parameter b: a) 0.60; b) 0.61; c) 0.62; d) 0.63; e) 0.64. In case (a) both PD lines are still continuos, in case (b) the NS line exists only for the period-16 cycle, (c-e) line of Neimark-Sacker bifurcation exist for both cycles.}
 \end{figure}

Further increase of $b$ leads to the formation of a new Q area, which can be regarded as the formation of the second rupture of the Feigenbaum line, which occurs at $0.6<b<0.61$. It emerges due to the appearance of NS bifurcation lines for all periodic orbits from the period-doubling cascade starting from the ones with higher periods: the NS bifurcation for period $2n$ appears at smaller value of parameter $b$ than for period $n$. In some interval of parameter $b$ NS bifurcations appear for all cycles down to period 2. The Feigenbaum line, of course, ruptures at the lower border of this interval of $b$. This evolution of the bifurcation lines with the change of $b$ is shown in Fig.\,{\ref{fignew}}.

\begin{figure*}
 \includegraphics[width=17.8cm]{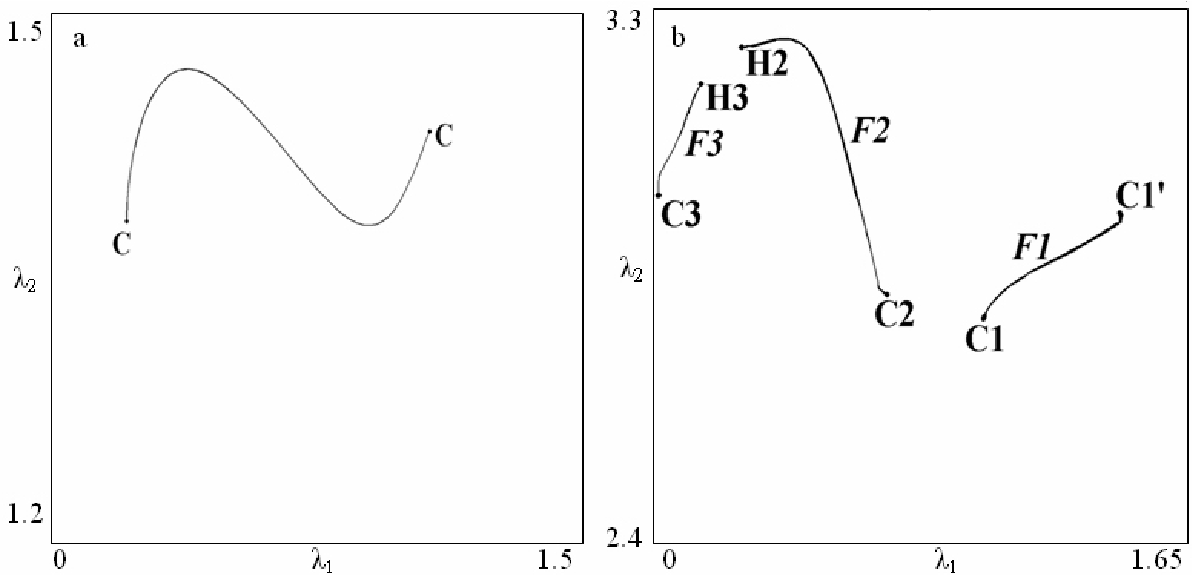}%
 \caption{\label{fig3} Sketch of the Feigenbaum critical line of Eq.\,(\ref{maineq}) at $b=0$ (a) and $b=0.7$ (b) (magnification of the parameter plane parts marked with black rectangles in Fig.\,{\ref{fig1}} and Fig.\,{\ref{fig2}}\,c). F1, F2, F3 denote fragments of the Feigenbaum line, C, C1, C1', C2, C3, H2, H3 --- critical points of C and H type correspondingly.}
 \end{figure*}

\begin{figure*}
 \includegraphics[width=17.8cm]{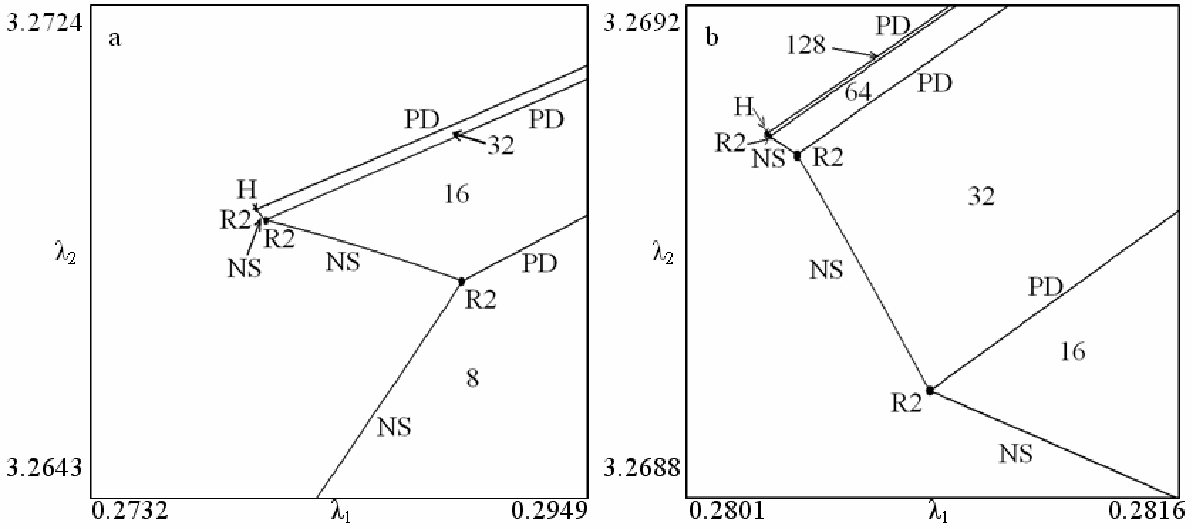}%
 \caption{\label{fig4}(color online) Structure of bifurcation lines of Eq.\,(\ref{maineq}) at $b=0.7$. PD denotes period-doubling line, NS --- line of Neimark-Sacker bifurcation, R2 --- resonance 1:2 point, H --- location of the critical point of H type, numbers denote the period of the stable regime in the corresponding area.}
 \end{figure*}

Fig.\,{\ref{fig3}} represents the structure of the Feigenbaum line for larger $b$ values ($b=0.7$ in Fig.\,{\ref{fig3}}\,b ) compared to the one for the coupled logistic maps at $b=0$ (Fig.\,{\ref{fig3}}\,a). We note that instead of critical points of C-type critical points of another type denoted by H1 and H2 emerge on both ends of this rupture. Fig.\,{\ref{fig4}} represents the picture of PD and NS lines at the right border of the Q area at $b=0.7$ (vicinity of the point H2 in Fig.\,{\ref{fig3}}\,b). One finds that all period-doubling lines terminate at the resonance 1:2 points, i.e. points in which one pair of multipliers have modulus less than 1 while the other two are equal to (-1,-1) \cite{kuzyubif}. Such sequence of points is typically the route to the critical point of Hamiltonian type, or H-point \cite{KuzStat, savin}. The coordinates of these terminal points are presented in Table\,\ref{tab2}. Again they accumulate to a certain limit, which we assume to be the critical point of H type. Using the corresponding scaling constant $\delta_H=8.72 \dots$ \cite{reichl} we obtain the expected location of this point at $\lambda_1=0.280283808, \lambda_2=3.269106587$ and calculate the multipliers of cycles of periods 32 and 64 in it. We obtain $\mu_1= �2.06$ and $\mu_2= �0.48$ which are rather close to the universal values $\mu_1= �2.0574783 \dots$ and $\mu_2= �0.4860318 \dots$ \cite{KuzStat}. We conclude that the critical point of H-type appears here as a terminal point of the Feigenbaum line. The dynamics on the other side of the Q area is the same, so that there exists another critical point of H type.
 \begin{table}
 \caption{\label{tab2}Coordinates of the resonance 1:2 points converging to the critical point of H type at $b=0.7$ (H2 point in notations of Fig.\,{\ref{fig3}}\,b)}
 \begin{tabular}{ccc}
\hline \hline
Period & $\lambda_1$ & $\lambda_2$ \\
\hline
16 & 0.28080895 & 3.26891802\\
32 & 0.28038063 & 3.26909124\\
64 & 0.28029262 & 3.26910445\\
128 & 0.28028497 & 3.26910634\\
256 & 0.28028394 & 3.26910656\\
\hline \hline
 \end{tabular}
 \end{table}

It is worth mentioning that we observe a change of the type of NS-bifurcation for different periodic orbits from the period-doubling cascade, e.g., it is supercritical for periods $n$ and $4n$ while subcritical for periods $2n$ and $8n$. The same behavior has been obtained in \cite{savin} for another system in a similar situation. Hence, it seems to be typical for the appearance of the H-type critical point in dissipative systems.

At both sides of the other rupture (left end of fragment F1 and right end of fragment F2 in Fig.\,{\ref{fig3}}\,b) the Feigenbaum line fragments terminate with C-type critical points, similar to the picture at smaller values of $b$. The coordinates of the ``fold-flip'' terminal points from the converging sequence illustrating this fact are presented in Table\,\ref{tab3}.
 \begin{table}
 \caption{\label{tab3}Coordinates of the ``fold-flip'' points at the left end of the Feigenbaum line rupture at $b=0.7$ (vicinity of C2 point in notations of Fig.\,{\ref{fig3}}\,b)}
 \begin{tabular}{ccc}
\hline \hline
Period & $\lambda_1$ & $\lambda_2$ \\
\hline
8 & 0.70190094 & 2.81854645\\
16 & 0.70244460 & 2.81438624\\
32 & 0.70827390 & 2.80831558\\
64 & 0.70759608 & 2.80900826\\
128 & 0.70847610 & 2.80802103\\
256 & 0.70817322 & 2.80835790\\
512 & 0.70836562 & 2.80814228\\
1024 & 0.70828259	 & 2.80823500\\
2048 & 0.70833027	 & 2.80818167\\
\hline \hline
 \end{tabular}
 \end{table}

Finally, we have three fragments of the Feigenbaum line which we call F1, F2 and F3 respectively going from right to left (see Fig.\,{\ref{fig3}}\,b). The F1 fragment terminates with critical points of C-type at both ends, while F2 and F3 possess critical points of different types (C and H respectively) at their two ends. This corresponds well with the conjecture about the possible terminal points of the Feigenbaum line which could be of C or H type in the general case \cite{Kuz1992}.

\section{\label{3} Formation of the rupture}

So far we have identified the critical behavior in the vicinity of the terminal points of fragments of the Feigenbaum lines for different levels of dissipation. But this does not answer the question how the first rupture of the Feigenbaum line is formed.

In order to reveal the mechanism of this rupture formation we have computed a number of charts of dynamical regimes for different values of dissipation $b$. First we recall the important property of chaotic dynamics, that it is interspersed with periodic windows of different periods  \cite{milnor88, graczyk, joglekar}. The skeletons of these regions of periodic dynamics within the chaotic region in the two-dimensional parameter space are the spring area and crossroad area structures \cite{Mira1, Mira2, Mira3}. When changing the dissipation parameter $b$ these periodic regions ``move'' through parameter space, i.e. they change their location as well as their size. It turns out that for a particular $b$ value one of the periodic regions collides with the main periodic region and merges subsequently with it.

\begin{figure*}
 \includegraphics[width=17.8cm]{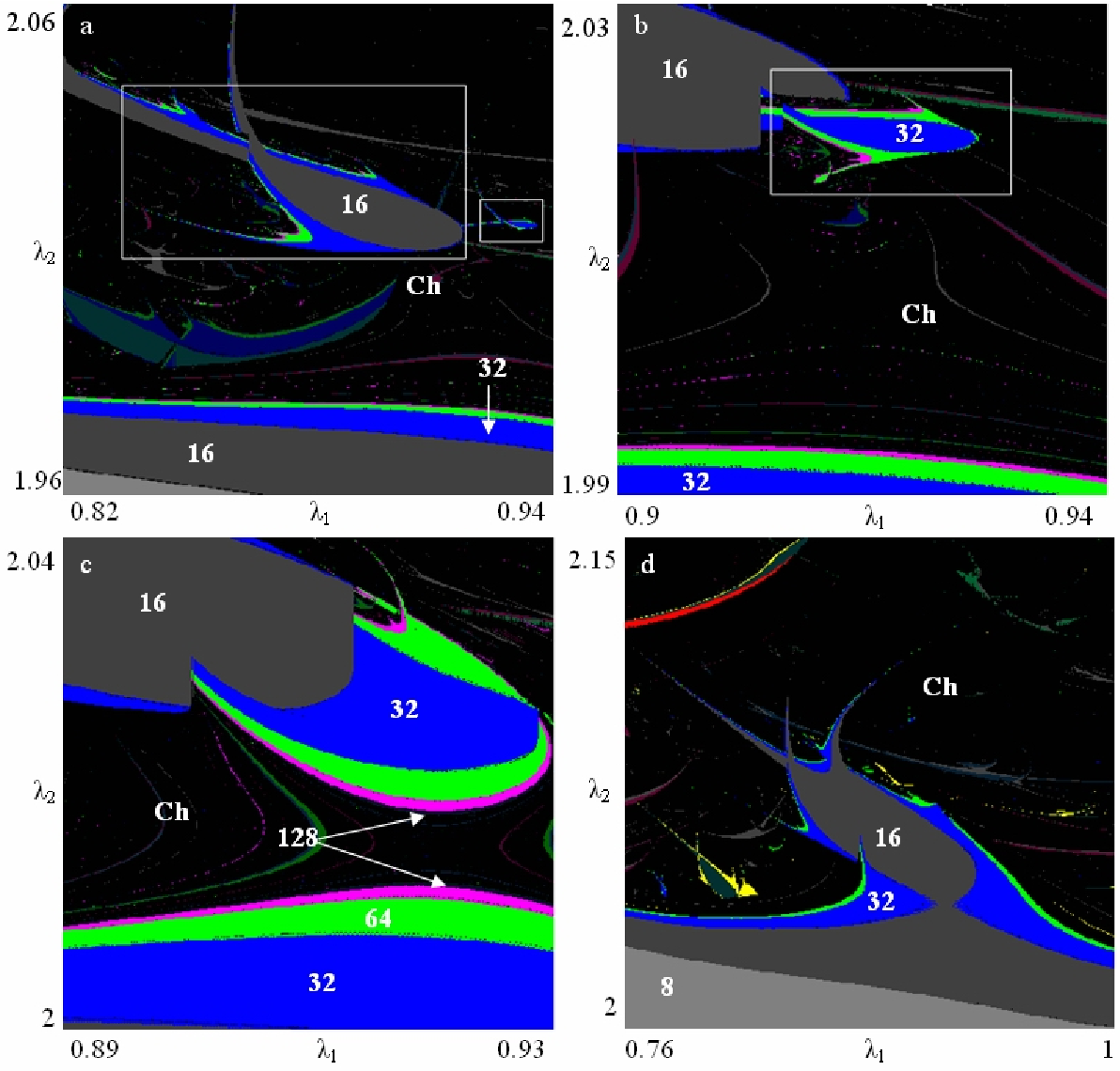}%
 \caption{\label{fig5}Fragments of the parameter plane illustrating the process of the formation of the Feigenbaum line rupture. $\varepsilon=0.4$, $b$ values: a)\,0.31, b)\,0.316, c)\,0.3215, d)\,0.334. Color code as in Fig.\,{\ref{fig1}}. For better representation the period-32 window is shown fully, while from the period-16 window one part is cut out, since this part would overlay the period-32 window. Both attractors, period-16 and period-32, coexist and possess different basins of attraction.}
 \end{figure*}

To illustrate this process we show several parts of the parameter plane at $\varepsilon=0.4$ and different $b$ values in Fig.\,{\ref{fig5}}. In Fig.\,{\ref{fig5}}\,a one can see the periodic window based on the spring area for the period-16 cycle together with the cascade of the periodic windows consisting of a series of spring areas for cycle of period 32, 64 etc. which are highlighted here and further in Fig.\,{\ref{fig5}}\,b by white rectangles. Figures\,{\ref{fig5}}\,b,\,c show the evolution of these periodic windows with increasing $b$ which results in a merging of the whole structure with the main periodic area (Fig.\,{\ref{fig5}}\,d).

It is necessary to recall here briefly the structure of a typical periodic window. Crossroad area and spring area structures based on the cycle of period $n$ are formed by two fold lines for this cycle emanating from a cusp point \cite{Mira1, Mira2, Mira3}, hence one can find multistability and the parameter space becomes divided into two ``multistability sheets'', which mean that in some region in the parameter space two attractors with different basins of attraction and independent dynamics coexist. On each of these multistability sheets there exists a period-doubling line. Since the periodic window usually consists of a cascade of such structures with periods $n, 2n$ etc., at each level of this period-doubling cascade a new splitting of the parameter space into multistability sheets occurs, and finally at the border of chaos one obtains a very complicated fractal-like structure with an infinite number of Feigenbaum line fragments \cite{KuzStat, KuzChua}. This complex hierarchical structure of bifurcation lines is located in small regions of the parameter space and is no at all related with the usual transition to chaos observed starting from the main periodic area. However, when the merging of this periodic window with the main periodic area occurs, this complicated bifurcation structure affects also the transition to chaos from the main periodic area. 

\begin{figure*}
 \includegraphics[width=17.8cm]{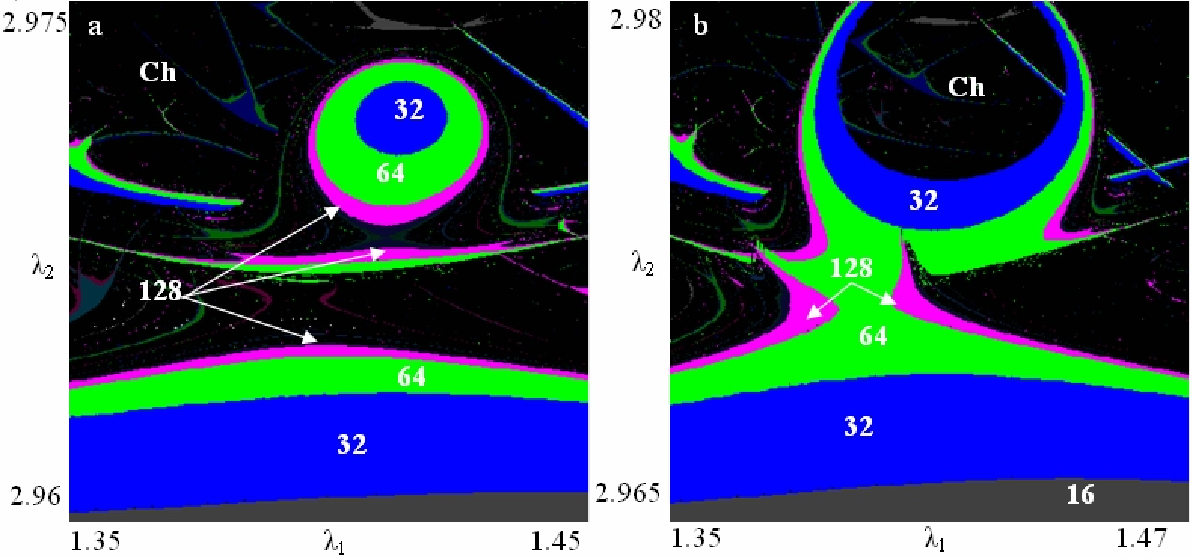}%
 \caption{\label{fig6}(color online) Fragments of the parameter plane illustrating the process of the formation of the Feigenbaum line rupture at $\varepsilon=0.2$. $b$ values: a)\,0.6558, b)\,0.6575. Color code as in Fig.\,{\ref{fig1}}.}
 \end{figure*}

\begin{figure*}
 \includegraphics[width=17.8cm]{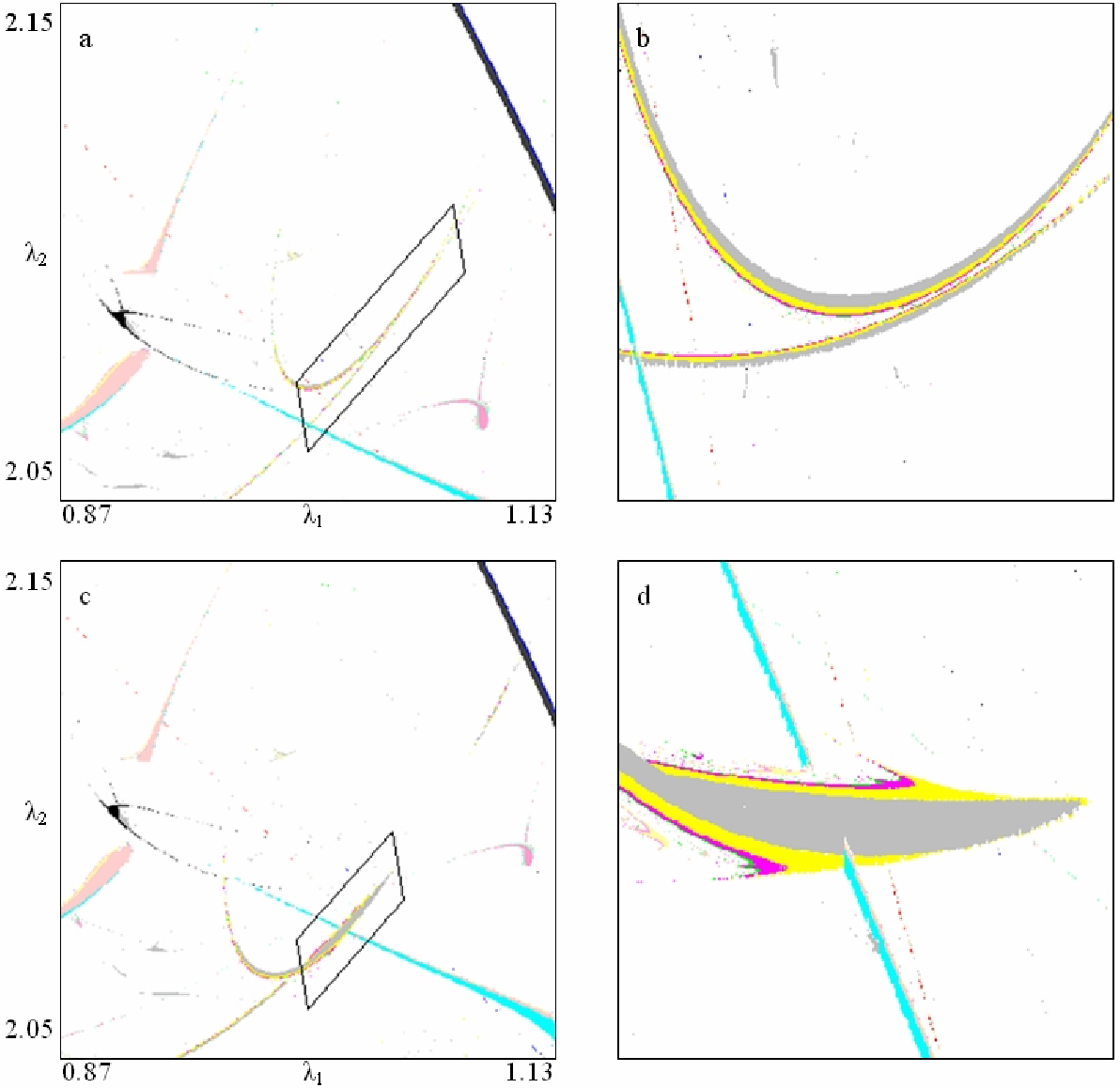}%
 \caption{\label{fig7}(color online) Fragments of the parameter plane illustrating the process of the collision of two periodic windows. Pictures in the right column are the enlargements of the black parallelograms from the left one. $\varepsilon=0.37$, $b$ values: a),\,b)\,0.298, c),\,d)\,0.304. Color code differs from used in Fig.\,{\ref{fig1}}: white --- chaotic region, grey etc. --- periodic windows, the main period of the concerning windows is 16.}
 \end{figure*}

Though the rupture of the Feigenbaum line has only been demonstrated for the coupling $\varepsilon=0.4$, the described mechanism of rupture formation exists in a certain interval of the coupling parameter $\varepsilon$. However, it is important to note that the definite shape of the periodic window involved in the process of rupture could vary. For example, for $\varepsilon=0.2$ the described process occurs not with a spring area structure but with a ``ring-shaped'' periodic window (see Fig.\,{\ref{fig6}}). Additionally, varying $\varepsilon$ and $b$ one can find a transition from one type of periodic window to another. As an example we show the collision of this ``ring-shaped'' structure with another periodic window based on the crossroad area structure in Fig.\,{\ref{fig7}}\,a,\,b, which results in the formation of two spring area based periodic windows (Fig.\,{\ref{fig7}}\,c,\,d). This could be interpreted as an indicator for the existence of a highly complicated connection between different periodic windows in the four-dimensional parameter space.

\section{\label{4} Summary}
In this paper we have studied two conservatively coupled H\'enon maps with a special focus on the change in the transition to chaos as the strength of dissipation is varied. In the limit of infinitely strong dissipation the system turns into two coupled logistic maps. For this system it is known that in the region where subsystems are sufficiently non-identical the transition to chaos appears via a period-doubling cascade accumulating at the Feigenbaum line which marks the transition to chaos. As dissipation is lowered more complicated dynamics arises which involves two ruptures of this Feigenbaum line. The first rupture yields two fragments which are terminated by critical points of C-type. The second one leading to existence of three fragments of the Feigenbaun line yields critical points of another type. This second rupture corresponds to the emergence of a region of quasiperiodic behavior in parameter space. The terminal points of the fragments of the Feigenbaum line facing the quasiperiodic parameter region are both of H-type. This means that those two fragments possess critical points of different type on each of its ends related to different scaling properties. This seems to be the typical situation for maps characterized by more than one parameter according to \cite{Kuz1992} but to the best of our knowledge our study is the first one demonstrating this phenomenon in numerical simulations.

Additionally we have demonstrated the mechanism of formation of the first rupture. Our study shows that the periodic windows, which are usually present in the chaotic parameter region are moving with changing parameter values and approaching the main periodic area. Finally one such periodic window merges with the main periodic area giving rise to the first rupture.

\begin{acknowledgments}
First of all, we want to thank Dr.\,Igor R.\,Sataev for his useful advices and discussions concerning critical behavior. D.\,V.\,S and A.\,V.\,S. would like to thank the Rusian Foundation for Basic Researches for financial support, projects 12-02-31089 and 14-02-31067. D.\,V.\,S. would also like to thank German Academic Exchange Service and the Direction of development of the National Research University ``Chernyshevsky Saratov State University'' for financial support of his visits in Oldenburg and Ulrike Feudel and her group of Complex Systems for their hospitality.
\end{acknowledgments}

\bibliography{final}

\end{document}